\documentclass{article}
\usepackage{times}
\usepackage{amsfonts}
\usepackage{graphicx}
\usepackage[pdfmark]{hyperref}
\begin{document}
\noindent
{\Large REMARKS ON THE REPRESENTATION THEORY OF THE MOYAL PLANE}
\vskip1cm
\noindent
{\bf J.M. Isidro}${}^{1,2,a}$, {\bf P. Fern\'andez de C\'ordoba}${}^{1,b}$,  {\bf J.M. Rivera--Rebolledo}$^{3, c}$\\
and {\bf J.L.G. Santander}${}^{4,d}$\\
${}^{1}$Instituto Universitario de Matem\'atica Pura y Aplicada,\\ Universidad Polit\'ecnica de Valencia, Valencia 46022, Spain\\
${}^{2}$Max--Planck--Institut f\"ur Gravitationsphysik, Albert--Einstein--Institut,\\ D--14476 Golm, Germany\\
${}^{3}$Escuela Superior de F\'{\i}sica y Matem\'aticas, Instituto Polit\'ecnico Nacional, \\
Departamento de F\'{\i}sica, U. P. Adolfo L\'opez Mateos, 07738 Lindavista,\\
M\'exico D. F., Mexico\\
${}^{4}$C\'atedra Energesis de Tecnolog\'{\i}a Interdisciplinar, Universidad Cat\'olica de Valencia,\\ C/ Guillem de Castro 94, Valencia 46003, Spain\\
${}^{a}${\tt joissan@mat.upv.es}, ${}^{b}${\tt pfernandez@mat.upv.es}, \\
${}^{c}${\tt jrivera@esfm.ipn.mx}, ${}^{d}${\tt jlgonzalez@mat.upv.es} \\
\vskip.5cm
\noindent
{\bf Abstract} We present an explicit construction of a unitary representation of the commutator algebra satisfied by position and momentum operators on the Moyal plane.

\section{Introduction}\label{einfuehrung}

There has been a lot of activity recently around physics on noncommutative spaces (for reviews see, {\it e.g.}, \cite{SZABO1, SZABO2} and refs. therein). Quite naturally, this has been accompanied by research into the foundations of quantum mechanics on noncommutative spaces. As was already the case in the early days of quantum mechanics, one key question is how to represent the symmetry algebra of the problem under consideration. The symmetry algebra encodes the kinematics, regardless of the dynamics. Classically the kinematics is summarised by the Poisson brackets $\{q_i,q_j\}=0$,  $\{p_i,p_j\}=0$, $\{q_i,p_j\}=\delta_{ij}$, while quantum--mechanically the latter become commutators for the operators $Q_i,P_j$, 
\begin{equation}
[Q_i,Q_j]=0, \qquad [P_i,P_j]=0, \qquad [Q_i,P_j]={\rm i}\hbar\delta_{ij}.
\label{dos}
\end{equation}
A celebrated theorem of Stone and von Neumann \cite{THIRRING} establishes that all unitary representations of the commutator algebra (\ref{dos}) are unitarily equivalent to that given by (complex infinite--dimensional, separable) Hilbert space, with position operators $Q_i$ acting multiplicatively on the wavefunctions and momentum operators $P_j$ acting by differentiation of the same wavefunctions.

When space becomes noncommutative, the commutators $[Q_i,Q_j]=0$ develop nonzero terms on their right--hand sides. For simplicity let us restrict our attention to the 2--dimensional case. We will also assume the simplest form of noncommutativity, namely that given by the Moyal plane: $[X,Y]={\rm i}\theta{\bf 1}$, with $\theta>0$. The corresponding symmetry algebra, that replaces eqn. (\ref{dos}) above, is expressed in eqn. (\ref{possonalg}) below. Then a natural question to ask is what becomes of the Stone--von Neumann theorem on the Moyal plane. Although this issue has been addressed in the literature \cite{SCHOLTZ1}, here we offer an alternative viewpoint. Specifically, we provide an explicit construction of a unitary representation of the symmetry algebra (\ref{possonalg}) on the Moyal plane, in terms of noncommutative oscillator modes. This representation will be used in an approach to quantum mechanics on the Moyal plane \cite{NOS}, an approach that has been demanded in the literature \cite{INDIO} and developed to some extent  \cite{INDIO2, INDIO3, INDIO4, INDIO5, SCHOLTZ2}. The idea underlying this approach is the following. Coordinates $X,Y$ on the Moyal plane are actually selfadjoint operators on Hilbert space. Now one expects quantum--mechanical wavefunctions to depend on the space coordinates. If the latter are operators, then wavefunctions too must be operators. This requires one to first identify a unitary representation on which $X,Y$, as selfadjoint operator--valued coordinates, act and satisfy the symmetry algebra (\ref{possonalg}). In this paper we tackle this problem, leaving the construction of operator--valued wavefunctions for a forthcoming publication \cite{NOS}.

\section{The noncommutative Poisson--Heisenberg algebra}\label{nonpo}

\subsection{The commutator algebra}\label{ssuno}

The noncommutative plane $\mathbb{R}_{\theta}^2$ is defined as the algebra of functions of two generators $X,Y$ satisfying the commutator $[X,Y]={\rm i}\theta{\bf 1}$, with $\theta>0$. We regard $\mathbb{R}_{\theta}^2$ as a two--dimensional configuration space endowed with noncommuting coordinates $X,Y$. On the corresponding noncommutative phase space $\mathbb{R}_{\theta, \hbar}^4$ we have the operators $X$, $Y$, $P_X$, $P_Y$ satisfying a commutator algebra that we postulate to be
\begin{equation}
[X,Y]={\rm i}\theta {\bf 1},\quad [X,P_X]=[Y,P_Y]={\rm i}\hbar{\bf 1},\quad [P_X,P_Y]=[X,P_Y]=[Y,P_X]=0,
\label{possonalg}
\end{equation}
We will call the set of eqns. (\ref{possonalg}) the  2--dimensional, {\it noncommutative Poisson--Heisenberg algebra}\/.  The time variable $t$ will be taken to commute with all generators $X,Y,P_X,P_Y$. We would like to observe that positing the above algebra amounts to positing the symplectic structure first derived in \cite{DUVAL}. One can  then show that choosing the standard (quadratic in momentum) Hamiltonian, the system will carry the so--called {\it exotic}\/ Galilean symmetry (for a recent review see \cite{SIGMA}).

It has been known for long that the {\it Bopp shift}
\begin{equation}
Y\mapsto Y-\frac{{\theta}}{\hbar}P_X
\label{vop}
\end{equation}
reduces the noncommutative Poisson--Heisenberg algebra (\ref{possonalg}) to the usual Poisson--Heisenberg algebra in two commuting space dimensions.

\subsection{Commutative oscillator modes}\label{com}

We will first construct a Hilbert--space representation for the commutator algebra (\ref{possonalg}), in terms of {\it commutative}\/ oscillator modes. This is of course trivial, but it will serve as a warmup exercise for the construction in terms of {\it noncommutative}\/ oscillator modes. Consider the usual harmonic oscillator eigenstates $\phi_n$ in 1 dimension,  where $n\in\mathbb{N}$. The space spanned by the $\phi_n$ is ${\ell}^2$, the Hilbert space of complex, square--summable sequences.
In two commuting dimensions $x,y$ we have the eigenstates $\phi_{nm}(x,y)=\phi_{n}(x)\phi_{m}(y)$. The latter form an orthonormal basis for the Hilbert space ${\ell}^2\times {\ell}^2$. Position and momentum operators $X',Y',P'_X,P'_Y$ can be defined on the space ${\ell}^2\times {\ell}^2$ as usual \cite{LANDAU}: acting on the first index,
\begin{equation}
X'\phi_{nm}:=\sqrt{\frac{\theta}{2}}\left(\sqrt{n+1}\,\phi_{n+1,m}+\sqrt{n}\,\phi_{n-1,m}\right),
\label{unoequis}
\end{equation}
\begin{equation}
P'_X\phi_{nm}:=\frac{{\rm i}\hbar}{\sqrt{2\theta}}\left(\sqrt{n+1}\,\phi_{n+1,m}-\sqrt{n}\,\phi_{n-1,m}\right).
\label{unopequis}
\end{equation}
{}For the second index we define the action of $Y',P'_Y$ similarly, with the sole difference that the (reverse) Bopp shift (\ref{vop}) must be taken into account:
\begin{equation}
Y'\phi_{nm}:=\sqrt{\frac{\theta}{2}}\left(\sqrt{m+1}\,\phi_{n,m+1}+\sqrt{m}\,\phi_{n,m-1}\right)+\frac{\theta}{\hbar}P'_X\phi_{nm},
\label{unoygriega}
\end{equation}
\begin{equation}
P'_Y\phi_{nm}:=\frac{{\rm i}\hbar}{\sqrt{2\theta}}\left(\sqrt{m+1}\,\phi_{n,m+1}-\sqrt{m}\,\phi_{n,m-1}\right).
\label{unopygriega}
\end{equation}
One verifies that the operators $X',Y',P'_X,P'_Y$ indeed satisfy the algebra (\ref{possonalg}).  We have denoted these operators with a prime because this representation is unsatisfactory for our purposes. Indeed, there is nothing noncommutative about the eigenstates $\phi_{nm}$: they are simply those of the harmonic oscillator on the commutative plane $\mathbb{R}^2$, noncommutativity being implemented in the algebra by means of the (inverse) Bopp shift. Instead one would like to have a representation space spanned by eigenstates $\psi_{nm}$ of the harmonic oscillator on the noncommutative plane $\mathbb{R}^2_{\theta}$. This will be done explicitly in section \ref{sstres}.

\subsection{Interlude}\label{ssdos}

Before moving on to noncommutative oscillator modes we need to recall some elementary facts \cite{THIRRING}. Consider the space $F$ of all entire functions $f:\mathbb{C}\rightarrow\mathbb{C}$ such that
\begin{equation}
f(z)=\sum_{n=0}^{\infty}\frac{c_n}{\sqrt{n!}}z^n, \qquad \sum_{n=0}^{\infty}\vert c_n\vert^2<\infty.
\label{bese}
\end{equation}
This space is Hilbert with respect to the scalar product
\begin{equation}
\langle f\vert \tilde f\rangle:=\frac{1}{2\pi{\rm i}}\int {\rm d}z^*\wedge {\rm d}z\, {f^*(z)}\tilde f(z){\rm e}^{-\vert z\vert^2},
\label{presc}
\end{equation}
where the asterisk denotes complex conjugation, and the integral extends over all $\mathbb{R}^2$ with $z=(x+{\rm i}y)/\sqrt{2}$. An orthonormal basis is given by the set of all complex monomials
\begin{equation}
f_n(z):=\frac{z^n}{\sqrt{n!}}, \qquad n\in\mathbb{N}.
\label{momo}
\end{equation}
The space $F$ is called {\it Bargman--Segal space}\/. The $f_n$ are in 1--to--1 correspondence with the harmonic oscillator eigenstates $\phi_n$ of section \ref{com}.

Next consider the following variant of Bargman--Segal space. Let us consider functions  $g:\mathbb{R}\rightarrow\mathbb{C}$  such that
\begin{equation}
g(x)=\sum_{n=0}^{\infty}\frac{c_n}{\sqrt{n!}}x^n, \qquad \sum_{n=0}^{\infty}\vert c_n\vert^2<\infty,
\label{besereal}
\end{equation}
the $c_n$ being complex coefficients. Here our functions $g$ are complex--valued analytic functions of one {\it real}\/ variable $x$. Call $G$ the space of all functions satisfying (\ref{besereal}). A basis for $G$ is given by the set of all real monomials
\begin{equation}
g_n(x):=\frac{x^n}{\sqrt{n!}}, \qquad n\in\mathbb{N}.
\label{basereal}
\end{equation}
We can define a scalar product on $G$ by declaring these monomials to be orthonormal,
\begin{equation}
\langle g_n\vert g_m\rangle :=\delta_{nm}, \qquad n,m\in\mathbb{N},
\label{toro}
\end{equation}
and extending the above to all elements of $G$ by complex linearity. This scalar product makes $G$ a complex Hilbert space. The difference with respect to Bargman--Segal space $F$ is that, the functions $g\in G$ depending on the real variable $x$ instead of the complex variable $z$, the scalar product on $G$ is no longer given by (\ref{presc}), nor by its real analogue. Indeed, given any two $g,\tilde g\in G$, the analogue of (\ref{presc}) for $G$ would be the integral
\begin{equation}
\int_{-\infty}^{\infty}{\rm d}x\,g^*(x)\tilde g(x){\rm e}^{- x^2}.
\label{noupre}
\end{equation}
Although this integral does define a scalar product on $G$, this scalar product does not make the basis (\ref{basereal}) orthogonal, as one readily verifies. Therefore  one, and only one, of the following properties can be satisfied:\\
{\it i)} the space $G$ is Hilbert with respect to the scalar product (\ref{noupre}), but the monomial basis (\ref{basereal}) is not orthogonal with respect to it;\\
{\it ii)} the space $G$ is Hilbert with respect to the scalar product (\ref{toro}), and the monomial basis (\ref{basereal}) is indeed orthonormal with respect to it, but this scalar product is not given by the integral (\ref{noupre}).\\
This being the case, we settle in favour of condition {\it ii)} above as our choice for the Hilbert space $G$.

{}Finally, the construction given by eqns.  (\ref{besereal})--(\ref{toro}) can be straightforwardly extended to complex--valued, analytic functions of {\it two}\/ real variables $x,y$. This will be used next.

\subsection{Noncommutative oscillator modes}\label{sstres}

Next we construct a unitary, Hilbert--space representation for the algebra (\ref{possonalg}), in terms of noncommutative oscillator modes.  It will be based on the Hilbert space, just mentioned in section \ref{ssdos}, of complex--valued, analytic functions of two real variables---but with {\it noncommuting, selfadjoint operators}\/ replacing the real variables.

Consider first an auxiliary copy ${\cal H}$ of the Heisenberg algebra, spanned by operators $V,W,{\bf 1}$ satisfying $[V,W]={\rm i}\theta {\bf 1}$, where both $V$ and $W$ have dimensions of length. The algebra ${\cal H}$ is realised in the standard way: $V$ acts on auxiliary wavefunctions $h(v)$ by multiplication, $Vh(v)=vh(v)$, and $W$ acts by differentiation, $Wh(v)=-{\rm i}\theta{\rm d}h/{\rm d}v$.  That the dimension of $\theta$ is length squared, rather than that of an action, should not bother us, since ${\cal H}$ is an auxiliary construct. The corresponding Hilbert space of the wavefunctions $h(v)$, also termed auxiliary, is $L^2(\mathbb{R}, {\rm d}v)$. This Hilbert space, however, is {\it not}\/ the carrier space of the unitary representation of the algebra (\ref{possonalg}) that we are looking for. To reiterate, the algebra $[V,W]={\rm i}\theta {\bf 1}$ just introduced, although  isomorphic to the subalgebra $[X,Y]={\rm i}\theta {\bf 1}$ contained in (\ref{possonalg}), acts on the auxiliary space $L^2(\mathbb{R}, {\rm d}v)$, while the space on which the algebra $[X,Y]={\rm i}\theta {\bf 1}$ will act is about to be defined below.

Next let $U({\cal H})$ denote the universal enveloping algebra of ${\cal H}$. By definition, $U({\cal H})$ is the algebra of  polynomials in the operators $V,W,{\bf 1}$, of arbitrarily high degree, with $V$ and $W$ satisfying $[V,W]={\rm i}\theta{\bf 1}$. Some suitable completion of $U({\cal H})$, denoted $\overline{U({\cal H})}$ and to be constructed presently, is the space of convergent power series in $V,W$. We take an arbitrary vector of $\overline{U({\cal H})}$ to be an expression of the form
\begin{equation}
\psi(V,W)=\sum_{n,m=0}^{\infty}\frac{c_{nm}}{\sqrt{n!m!\,\theta^{n+m}}}V^nW^m,
\label{funzia}
\end{equation}
where the $c_{nm}$ are complex coefficients, such that the above series converges (in a sense to be specified presently). The factor $(\theta^{n+m})^{-1/2}$ ensures that all summands are dimensionless.  From now we will prescribe all vectors of $\overline{U({\cal H})}$  to be normal--ordered, {\it i.e.}, $V$ will {\it always}\/ be assumed to precede $W$, if necessary by applying the commutator $[V,W]={\rm i}\theta{\bf 1}$.

A basis for $\overline{U({\cal H})}$ is given by the vectors
\begin{equation}
\psi_{nm}(V,W)= \frac{1}{\sqrt{n!m!\,\theta^{n+m}}}V^{n}W^{m}, \qquad  n,m\in\mathbb{N}.
\label{sonrio}
\end{equation}
The simplest choice for a scalar product on $\overline{U({\cal H})}$ is to declare the basis vectors (\ref{sonrio}) orthonormal,
\begin{equation}
\langle\psi_{n_1m_1}\vert\psi_{n_2m_2}\rangle:=\delta_{n_1n_2}\delta_{m_1m_2},
\label{skalar}
\end{equation}
and to extend (\ref{skalar}) to all of $\overline{U({\cal H})}$ by complex linearity. Then the squared norm of the vector (\ref{funzia}) equals $\sum_{nm}\vert c_{nm}\vert^2$:
\begin{equation}
\vert\vert\psi(V,W)\vert\vert^2=\sum_{n,m=0}^{\infty}\vert c_{nm}\vert^2.
\label{norma}
\end{equation}
Since this norm must be finite, this identifies $\overline{U({\cal H})}$ as the Hilbert space of square--summable complex sequences $\left\{c_{nm}\right\}$ in two indices $n,m$, {\it the latter taken to be normal--ordered}\/ as in (\ref{sonrio}); this defines the completion of $U({\cal H})$ referred to above.
It is worthwhile to observe that, although the vectors (\ref{funzia}) are unbounded operators in their action on the auxiliary Hilbert space $L^2(\mathbb{R}, {\rm d}v)$, the same vectors {\it do}\/ have a finite norm as elements of the Hilbert space $\overline{U({\cal H})}$. This is so because the norm of $\psi(V,W)$ in (\ref{norma}) is being measured by means of the complex coefficients $c_{nm}$, not by means of the operator norms of $V,W$ (themselves infinite). We will henceforth call the $\psi_{nm}$ of (\ref{sonrio}) {\it noncommutative oscillator modes}\/.

The Hilbert space $\overline{U({\cal H})}$ just constructed will become the carrier space of a representation of the algebra  (\ref{possonalg}). For this we need to define the action of the operators $X,Y,P_X,P_Y$ on the noncommutative oscillator modes (\ref{sonrio}). We set
\begin{equation}
X\psi_{nm}:=\sqrt{\frac{\theta}{2}}\left(\sqrt{n+1}\,\psi_{n+1,m}+\sqrt{n}\,\psi_{n-1,m}\right)
\label{equis}
\end{equation}
and
\begin{equation}
P_X\psi_{nm}:=\frac{{\rm i}\hbar}{\sqrt{2\theta}}\left(\sqrt{n+1}\,\psi_{n+1,m}-\sqrt{n}\,\psi_{n-1,m}\right).
\label{pequis}
\end{equation}
{}For the second index we define the action of $Y,P_Y$ similarly, with the sole difference that the (reverse) Bopp shift (\ref{vop}) must be taken into account:
\begin{equation}
Y\psi_{nm}:=\sqrt{\frac{\theta}{2}}\left(\sqrt{m+1}\,\psi_{n,m+1}+\sqrt{m}\,\psi_{n,m-1}\right)+\frac{\theta}{\hbar}P_X\psi_{nm}
\label{ygriega}
\end{equation}
and
\begin{equation}
P_Y\psi_{nm}:=\frac{{\rm i}\hbar}{\sqrt{2\theta}}\left(\sqrt{m+1}\,\psi_{n,m+1}-\sqrt{m}\,\psi_{n,m-1}\right).
\label{pygriega}
\end{equation}
{}Finally, the operators $X,Y,P_X,P_Y$ so defined are Hermitian and  satisfy the algebra (\ref{possonalg}) as desired. The above $X,Y,P_X,P_Y$ are distinguished notationally from the operators $X',Y',P'_X,P'_Y$ of (\ref{unoequis})--(\ref{unopygriega}) in order to stress the fact that they are actually different operators acting on different spaces\footnote{All infinite--dimensional, complex, separable Hilbert spaces being unitarily isomorphic, the above statement is to be understood as {\it different realisations of Hilbert space}.}, even if the two sets of operators satisfy the same algebra (\ref{possonalg}). 

It is worth pointing out that the oscillator representation given above bears a close similarity with the one constructed in \cite{PETER} from the nonrelativistic limit of anyons.

\section{Higher dimensions}\label{alton}

The previous results can be easily generalised to higher--dimensional Moyal spaces. Let us outline the main results in dimension 3. Here the noncommutative Poisson--Heisenberg algebra reads 
\begin{equation}
[X,Y]= [Y,Z]= [Z,X]={\rm i}\theta{\bf 1},\qquad [X,P_X]=[Y,P_Y]=[Z,P_Z]={\rm i}\hbar{\bf 1},
\label{cogebr}
\end{equation}
all other commutators vanishing identically. For simplicity we have assumed the generators so normalised that there is just one independent noncommutativity parameter $\theta$. The Bopp shift that reduces (\ref{cogebr}) to the standard Poisson--Heisenberg algebra is
\begin{equation}
Y\mapsto Y-\frac{\theta}{\hbar}P_X, \qquad Z\mapsto Z+\frac{\theta}{\hbar}P_X-\frac{\theta}{\hbar}P_Y.
\label{horst}
\end{equation}
The algebra (\ref{cogebr}) contains three independent copies of the noncommutative plane. Therefore we will need three auxiliary copies ${\cal H}_1$, ${\cal H}_2$ and ${\cal H}_3$ of the Heisenberg algebra of section \ref{sstres}, the $j$--th copy having generators $V_j$, $W_j$ satisfying
\begin{equation}
[V_j,W_j]={\rm i}\theta{\bf 1},\qquad j=1,2,3.
\label{koehler}
\end{equation}
We convene to associate the index values $j=1,2,3$ with the respective commutators $[X,Y]={\rm i}\theta{\bf 1}$,  $[Y,Z]={\rm i}\theta{\bf 1}$ and $[Z,X]={\rm i}\theta{\bf 1}$, each one of which spans a copy of the Moyal plane. Following (\ref{sonrio}), the noncommutative oscillator modes $\psi_{n_jm_j}$ corresponding to the $j$--th copy of the Moyal plane are
\begin{equation}
\psi_{n_jm_j}:=\frac{1}{\sqrt{n_j!m_j!\,\theta^{n_j+m_j}}}V_j^{n_j}W_j^{m_y}, \quad n_j,m_j\in\mathbb{N},\quad j=1,2,3.
\label{ncosmm}
\end{equation}
The above $\psi_{n_jm_j}$ provide a complete orthonormal set for the space $\overline{U({\cal H}_j)}$. Position and momentum operators $X^{(1)},Y^{(1)}, P_{X^{(1)}},P_{Y^{(1)}}$ can be defined on $\overline{U({\cal H}_1)}$ as in section \ref{sstres}; by the same token we define the action of $Y^{(2)},Z^{(2)},P_{Y^{(2)}},
P_{Z^{(2)}}$ on  $\overline{U({\cal H}_2)}$, and the action of $Z^{(3)},X^{(3)},P^{(3)}_Z,P^{(3)}_X$ on $\overline{U({\cal H}_3)}$. 

Next consider the tensor product space
\begin{equation}
\overline{U({\cal H}_1)}\otimes \overline{U({\cal H}_2)}\otimes \overline{U({\cal H}_3)}. 
\label{parrrf}
\end{equation}
A complete orthonormal set on this product space is given by the tensor product states 
\begin{equation}
\psi_{{\bf n}{\bf m}}:=\psi_{n_1m_2}\otimes\psi_{n_2m_2}\otimes\psi_{n_3m_3}, \qquad {\bf n},{\bf m}\in\mathbb{N}^3,
\label{ttpnn}
\end{equation}
where ${\bf n}=(n_1,n_2,n_3)$ and ${\bf m}=(m_1,m_2,m_3)$. We can extend the above position and momentum operators to act on all of $\overline{U({\cal H}_1)}\otimes \overline{U({\cal H}_2)}\otimes \overline{U({\cal H}_3)}$ in the obvious way. Namely, $X^{(1)}$ on $\overline{U({\cal H}_1)}$ is extended as
$
X^{(1)}\mapsto X^{(1)}\otimes{\bf 1}_2\otimes{\bf 1}_3,
$
while  $Y^{(2)}$ on $\overline{U({\cal H}_2)}$ is extended as
$
Y^{(2)}\mapsto {\bf 1}_1\otimes Y^{(2)}\otimes{\bf 1}_3, 
$
and so forth. In this way all 3 sets of position and momentum operators 
\begin{equation}
X^{(1)},Y^{(1)}, P^{(1)}_X,P^{(1)}_Y, \qquad Y^{(2)},Z^{(2)},P^{(2)}_Y,P^{(2)}_Z, \qquad Z^{(3)},X^{(3)},P^{(3)}_Z,P^{(3)}_X,
\label{allthree}
\end{equation}
each one of them initially defined to act only on the correponding space $\overline{U({\cal H}_j)}$, are now defined on the tensor product space (\ref{parrrf}). We will not distinguish notationally between the operators (\ref{allthree}) and their extensions to the tensor product space (\ref{parrrf}). On the latter we finally define
$$
X:=\frac{1}{\sqrt{2}}\left(X^{(1)}+X^{(3)}\right), \quad P_X:=\frac{1}{\sqrt{2}}\left(P^{(1)}_X+P^{(3)}_X\right)
$$
$$
Y:=\frac{1}{\sqrt{2}}\left(Y^{(1)}+Y^{(2)}\right), \quad P_Y:=\frac{1}{\sqrt{2}}\left(P^{(1)}_Y+P^{(2)}_Y\right)
$$
\begin{equation}
Z:=\frac{1}{\sqrt{2}}\left(Z^{(2)}+Z^{(3)}\right), \quad P_Z:=\frac{1}{\sqrt{2}}\left(P^{(2)}_Z+P^{(3)}_Z\right).
\label{longlast}
\end{equation}
One verifies that the operators (\ref{longlast}), acting on the tensor product states (\ref{ttpnn}), indeed satisfy the algebra (\ref{cogebr}) as desired.

We would like to remark that the commutator relations (\ref{cogebr}) break rotation invariance. However the latter can be restored if the constant parameter $\theta$ is promoted to a vector--valued function of the momentum, whenever this function is divergence--free in momentum space \cite{SIGMA, BERARD}.

\section{Discussion}\label{diskku}

In this paper we have constructed a unitary representation for the symmetry algebra (\ref{possonalg}). This latter algebra encodes the kinematics of quantum mechanics on the Moyal plane, regardless of whatever specific dynamics one wishes to consider. This representation has noncommutative oscillator modes as its building blocks. Such oscillator modes are the noncommutative analogues of ordinary harmonic oscillators on the commutative plane $\mathbb{R}^2$. We have also sketched how to generalise these noncommutative oscillator modes to higher dimensions, provided that the space noncommutativity is always of the Heisenberg--algebra type.

As stated in the introduction, the purpose of these noncommutative oscillator modes is to use them in a novel approach to quantum mechanics on noncommutative spaces. This approach does not make use of c--number valued wavefunctions that are multiplied together by means of a star--product. Rather, one looks for operator--valued wavefunctions already from the start \cite{INDIO}. Once space coordinates are operators, since wavefunctions will be functions of the coordinates, wavefunctions themselves will become operator--valued. The analysis carried out here is a necessary first step towards that goal. 

Looking beyond, a feature of emergent phenomena is that they arise as some form of coarse--grained, or thermodynamical, description of some microscopic physics that one does not have complete control of  \cite{CARROLL0}.  This much is true in general, and also of quantum mechanics in particular \cite{ELZE1, ELZE2, ELZE3, MATONE}, even before introducing noncommutativity. Now space noncommutativity also introduces a form of coarse graining, due to the existence of the quantum of area $\theta$---not on phase space, but on configuration space. In this sense, noncommutative quantum mechanics also falls within the category of emergent physics.

\vskip.5cm
\noindent
{\bf Acknowledgements}  We would like to thank the referee for constructive suggestions. J.M.I. thanks Max--Planck--Institut f\"ur Gravitationsphysik, Albert--Einstein--Institut (Golm, Germany), for hospitality. This work has been supported by Universidad Polit\'ecnica de Valencia under grant PAID-06-09, and by Generalitat Valenciana (Spain).\\

\end{document}